\documentstyle[emulateapj,pstricks]{article}

\def\Mpc{\mbox{Mpc}}
\def\mpch{\mbox{$h^{-1}$Mpc}}
\def\Mpch{\mbox{$h^{-1}$Mpc}}
\def\hMsun{h^{-1}{\ }{\rm M_{\odot}}}
\def\hMpc{h^{-1}{\ }{\rm Mpc}}
\def\hkpc{h^{-1}{\ }{\rm kpc}}

\def\Msun{\mbox{$M_\odot$}}
\def\msunh{\mbox{$h^{-1}$M$_\odot$}}
\def\Msunh{\mbox{$h^{-1}$M$_\odot$}}

\def\kms{\mbox{{\rm km s}$^{-1}$}}

\def\vc{\mbox{$V_{\rm circ}$}}
\def\Vc{\mbox{$V_{\rm circ}$}}
\def\etal{et al.}
\def\nstep{\mbox{$N_{\rm steps}$}}
\def\LCDM{\char'3CDM}
\def\Colin{Col{\'{\i}}n}

\def\mathnew{\mathsurround=0pt}
\def\ref{\par\noindent\hangindent=2pc \hangafter=1 }
\def\simov#1#2{\lower .5pt\vbox{\baselineskip0pt
    \lineskip-.5pt\ialign{$\mathnew#1\hfil##\hfil$\crcr#2\crcr\sim\crcr}}}

\def\'#1{\ifx#1i{\accent"13\i}\else{\accent"13#1}\fi}

\begin{document}
\submitted{submitted to the Astrophysical Journal}
\lefthead{WHERE ARE GALACTIC SATELLITES?}
\righthead{KLYPIN ET AL.}

\title{Where are the missing galactic satellites?}

\author{Anatoly A. Klypin, Andrey V. Kravtsov, and Octavio Valenzuela}
\affil{Astronomy Department, New Mexico State University, Box 30001, Dept.
4500, Las Cruces, NM 88003-0001}
\vspace{1mm}
\author{Francisco Prada}\vspace{1mm}

\affil{Instituto de Astronomia, Apartado Postal 877, 22900 Ensenada, Mexico}

\begin{abstract}
  
  According to the hierarchical clustering scenario, galaxies are
  assembled by merging and accretion of numerous satellites of
  different sizes and masses. This ongoing process is not 100\%
  efficient in destroying all of the accreted satellites, as evidenced
  by the satellites of our Galaxy and of M31. Using published data,
  we have compiled the circular velocity ($\Vc$) distribution function
  (VDF) of galaxy satellites in the Local Group. We find that within
  the volumes of radius of 570~kpc ($400~\hkpc$~ assuming the Hubble
  constant\footnote{Assuming $H_0=100h{\ }{\rm km{\ }s^{-1} Mpc^{-1}}$.}
  $h=0.7$) centered on the Milky Way and Andromeda, the average VDF is
  roughly approximated as $n(>\Vc) \approx 45(\Vc/10~\kms)^{-1}h^3{\rm
    Mpc}^{-3}$ for $\Vc$ in the range $\approx 10-70{\ }\kms$. The
  observed VDF is compared with results of high-resolution cosmological
  simulations.  We find that the VDF in models is very different from
  the observed one: $n(>\Vc) \approx 1200(\Vc/10~\kms)^{-2.75}h^3{\rm
    Mpc}^{-3}$. Cosmological models thus predict that a halo of the
  size of our Galaxy should have about 50 dark matter satellites with
  circular velocity $>20~\kms$ and mass $>3\times 10^8 \Msun$ within
  a 570~kpc radius. This number is significantly higher than the
  approximate dozen satellites actually observed around our Galaxy. 
  The difference is even larger if we consider abundance of satellites 
  in simulated galaxy
  groups similar to the Local Group. The models predict $\sim 300$
  satellites inside a $1.5{\ }\Mpc$ radius, while there only $\sim 40$
  satellites are observed in the Local Group. The observed and predicted
  VDFs cross at $\approx 50\kms$, indicating that the predicted abundance
  of satellites with $\Vc\gtrsim 50{\ }{\kms}$ is in reasonably good
  agreement with observations.
 
  We conclude, therefore, that unless a large fraction of the Local
  Group satellites has been missed in observations, there is a dramatic
  discrepancy between observations and hierarchical models, regardless
  of the model parameters. We discuss several possible explanations for
  this discrepancy including identification of some satellites with the
  High Velocity Clouds observed in the Local Group, and the existence
  of dark satellites that failed to accrete gas and form stars due
  either to the expulsion of gas in the supernovae-driven winds or to
  gas heating by the intergalactic ionizing background.

\end{abstract}


\section{Introduction}


Satellites of galaxies are important probes of the dynamics and masses of
galaxies. Currently, analysis of satellite dynamics is one of the best
methods of estimating the masses within large radii of our Galaxy and of
the Local Group (e.g., Einasto \& Lynden-Bell 1982; Lynden-Bell et al.
1983; Zaritsky et al. 1989; Fich \& Tremaine 1991), as well as the
masses of other galaxies (Zaritsky \& White 1994; Zaritsky et al.
1997).  Although the satellites of the Milky Way and Andromeda galaxy
have been studied for a long period of time, their number is still
uncertain.  More and more satellites are being discovered (Irwin et al.
1990; Whiting et al. 1997; Armandroff et al. 1998; Karachentseva \&
Karachentsev 1998) with a wide range of properties;
some of them are relatively large and luminous and have appreciable star
formation rates (e.g., M33 and the Large Magellanic Cloud; LMC). Exempting
the strange case of IC10, which exhibits a high star formation rate
($0.7M_{\odot}{\rm yr}^{-1}$; Mateo 1998), most of the satellites
are dwarf spheroidals and dwarf ellipticals with signs of only mild
star-formation of $10^{-3}M_{\odot}{\rm yr}^{-1}$. The star formation
history of the satellites shows remarkable diversity: almost every
galaxy is a special case (Grebel 1998; Mateo 1998). This diversity
makes it very difficult to come up with a simple general model for formation
of satellites in the Local Group.  Because of the generally low star formation
rates, it is not unexpected that the metallicities of the satellites
are low: from $\approx 10^{-2}$ for Draco and And III to $\approx
10^{-1}$ for NGC 205 and Pegasus (Mateo 1998). There are indications
that properties of the satellites correlate with their distance to the
Milky Way (MW) or Andromeda, with dwarf spheroidals and dwarf
ellipticals being closer to the central galaxy (Grebel 1997). Overall,
about 40 satellites in the Local Group have been found.

Formation and evolution of galaxy satellites is still an open problem.
According to the hierarchical scenario, small dark matter (DM) halos
should on average collapse earlier than larger ones. To some degree,
this is supported by observations of rotation curves of dark-matter
dominated dwarfs and low-surface-brightness galaxies. The curves
indicate that the smaller the maximum circular velocity, the higher
the central density of these galaxies. This is expected from the
hierarchical models in which the smaller galaxies collapse earlier when
the density of the Universe was higher (Kravtsov et al. 1998; Kormendy
\& Freeman 1998).  Thus, it is likely that the satellites of the MW
galaxy were formed before the main body of the MW was assembled.
Some of the satellites may have survived the very process of the MW
formation, whereas others may have been accreted by the MW or by the Local
Group at later stages. Indeed this sequence forms the basis of the
currently popular semi-analytical models of galaxy formation (e.g.,
Somerville \& Primack 1998, and references therein).

There have been a number of efforts to use the Local Group as a
cosmological probe. Peebles et al. (1989) modeled formation of the
Local Group by gravitational accretion of matter onto two seed masses.
Kroeker \& Carlberg (1991) found pairs of ``galaxies'' in cosmological
simulations and used them to estimate the accuracy of traditional mass
estimates.  Governato et al. (1997) studied the velocity field around
Local Group candidates in different cosmological models and Blitz \etal
(1998) simulated a group of galaxies and compared their results with
the observations of the high-velocity clouds in the Local Group.

Nevertheless, despite significant effort, theoretical predictions
of the abundance and properties of the satellites are far from been
complete. One of the difficulties is the survival of satellites inside
halos of large galaxies. This numerically challenging problem requires 
very high-resolution simulations in a cosmological context and has 
been addressed in different ways.  In the classical
approach (e.g., Lin \& Lynden-Bell
1982; Kuhn 1993; Johnston et al. 1995), one assumes a realistic
potential for the MW, a density profile for the satellites (usually
an isothermal model with a central core), and numerically follows a
satellite as it orbits around the host galaxy.  This approach lends many
valuable insights into the physical processes operating on
the satellites and alleviates some of the numerical problems. It
lacks, however, one important feature: connection with the
cosmological background. The host galaxy is implicitly assumed to be
stable over many billions of years which may not be realistic for a
typical galaxy formed hierarchically.  Moreover, the assumed isothermal
density profile of the satellite is different from profiles of typical
dark matter halos formed in hierarchical models (Navarro, Frenk \&
White 1997). Last but not least, the abundances of the satellites
can only be predicted if the formation of the satellites and of the
parent galaxy is modelled self-consistently. Thus, more realistic
cosmological simulations are necessary.

Unfortunately, until recently numerical limitations prevented the usage of
cosmological simulations to address satellite dynamics. Namely,
dissipationless simulations seemed to indicate that DM halos do not
survive once they fall into larger halos (e.g., White 1976; van Kampen
1995; Summers et al. 1995). It appears, however, that the premature
destruction of the DM satellites inside the virial radius of larger halos
was largely due to numerical effects (Moore, Katz, Lake 1996; Klypin et
al.  1997 (KGKK)). Indeed, recent high-resolution simulations show that
hundreds of galaxy-size DM halos do survive in clusters of galaxies
(KGKK; Ghigna et al. 1997; {\Colin} et al. 1998). Encouraged by this
success, we have undertaken a study of  the survival of satellites in
galaxy-size halos.

Dynamically, galactic halos are different from cluster-size halos
(mass $\gtrsim 10^{14}\hMsun$).  Clusters of galaxies are relatively
young systems in which most of the satellite halos have had time to
make only a few orbits. Galaxies are on average significantly older,
enabling at least some of their satellites to orbit for many dynamical
times.  This increases the likelihood of the satellite being destroyed
either from numerical effects of the simulation or the real processes
of dynamical friction and tidal stripping. The destruction of the 
satellites is, of course, counteracted by accretion of the new
satellites in an ongoing process of galaxy formation.  It is clear,
therefore, that to predict the abundances and properties of galactic
satellites, one needs to model self-consistently both the orbital dynamics
of the satellites and the formation process of the parent halo in a
cosmological setting. In this paper we present results from a study of
galactic satellite abundances in high-resolution simulations of two
popular variants of the Cold Dark Matter (CDM) models. As will be
described below, the dissipationless simulations used in our study are
large enough to encompass a cosmologically significant volume and, at
the same time, have sufficient resolution to make the numerical
effects negligible.

The paper is organized as follows.  In \S 2 we present the data that we
use to estimate the {\em observed} velocity function of satellites of
our Galaxy and M31. Cosmological simulations are presented and
discussed in \S 3. We compare the predicted and observed velocity
functions in \S 4 to show that the models predict considerably more lower
mass satellites than is actually observed in the Local Group. In \S 5
and 6 we discuss possible interpretation and implications of our
results and summarize our conclusions.

\begin{deluxetable}{lllccrcccc}
\tablecolumns{9}
\tablecaption{Satellites of the MW and Andromeda}
\tablehead{
\colhead{\vc}\hfil & \colhead{Milky Way } &\colhead{Andromeda } &
\colhead{Average} & \colhead{Comments}\\
(km/s) & $286/571$kpc & $286/571$kpc &  $286/571$kpc}
\startdata
10& 11 /13& 13 /15 & 12/14& Sculptor Carina Sextans LeoII
\nl 
& &  & & AndI-III,V,VI, CAS Pegasus\nl 
15&  7 /9 &  7 /8  &  7/8.5 & Phoenix Fornax LeoI UrsMin Draco Sagit Lgs3      \nl 
20&  2 /3 &  6 /7  &  4/5 &   IC1613    \nl 
30&  2 /3 &  6 /6  &  4/4.5 & SMC NGC6822 IC10 NGC147 NGC185       \nl 
50&  1 /1 &  3 /3  &  2/2 &  LMC      \nl 
70&  0 /0 &  3 /3  &  1.5/1.5&  M33 M32 NGC205     \nl 
\enddata
\end{deluxetable}


\section{Satellites in The Local Group}


There are about 40 known galaxies in the Local Group (Mateo 1998). Most
of them are dwarf galaxies with absolute magnitudes of $M_V\approx
-10-15$. While more and more galaxies are being discovered, most of the
new galaxies are very small and faint making it seem unlikely that too many
larger satellites have been missed. Therefore, we have decided to
consider only relatively massive satellites with estimated rotational
velocity or three-dimensional velocity dispersion of stars larger than
10~km/s.  In order to simplify the situation even further, we estimate
the number of satellites {\em per central galaxy}. There is a number of
arguments why this is reasonable. First, it makes comparison with
cosmological models much more straightforward.  This is justified to
some degree by the fact that the satellites in the Local Group cluster
around either the MW or M31 and there are only a few very remote ones of
unclear association with a central galaxy. We also believe that the
estimate of the satellite abundance per galaxy is more accurate because
it is relatively straightforward to find the volume of the sample,
which would be more difficult if we were to deal with the Local Group
as a whole\footnote{One of the problems would be choice of the outer
  boundary of the sample volume.}.

Using published results (Mateo 1998), we have compiled a list of
satellites of the Milky Way and of the M31 with estimated circular
velocities above the threshold of $10{\ }{\rm km/s}$. In our estimate of
abundances, we have not attempted to decide whether a satellite is
bound to its central galaxy or not. Satellites have been simply counted
if they lie within a certain radius from the center of their parent
galaxy.  We have chosen two radii to make the counts. The counts of DM
satellites were made for the same radii. The radii were chosen rather
arbitrarily to be $200\hkpc$ and $400\hkpc$. For a Hubble constant of
$h=0.7$ ($H_0=100h$~km/s/Mpc), which was assumed for our most realistic
cosmological model and which is consistent with current observational
results, the radii are 286~kpc and 571~kpc. The smaller radius is close
to a typical virial radius of a Milky Way size halo in our simulations.
The larger radius allows us to probe larger volumes (and, thus, gives
better statistics) both in simulations and in observations.
Unfortunately, observational data may become less complete for this
radius.

Since we cannot estimate the luminosities of galaxies associated with
DM satellites in dissipationless simulations, we have chosen circular
velocity \Vc~ to characterize both the dark halos and the satellite
galaxies. The circular velocity can be estimated for galaxies (with
some uncertainties) and for the DM halos. For spiral and irregular
galaxies we used the rotational velocity, which is usually measured
from 21~cm HI observations. For ellipticals and dwarf spheroidals we
used observed line-of-sight velocity dispersion of stars, which was
multiplied by $\sqrt{3}$ to give an estimate of \Vc. Using our
numerical simulations we confirmed that this gives a reasonably
accurate estimate of \Vc~ with an error less than $\sim 10\%-20\%$.
Table 1 lists the number of satellites with \Vc~ larger than a given
value (first column) for the Milky Way galaxy (second column) and M31
(third column). The forth column gives the average number of
satellites and the fifth column lists names of the satellites in given
velocity bin. Figures 4 and 5, discussed in detail below, present the
cumulative circular velocity distribution of the observed satellites
in MW and M31 within 286~kpc and 571~kpc radius from the central
galaxies.

A few special cases should be mentioned. There are no measurements of
velocity dispersion for AND I-III and the other two satellites of M31, AND
V and VI, do not have measured magnitudes. Given that they all seem to
have the properties of a dwarf spheroidal, we think it is reasonable to
expect that they have \Vc~ in the range 10-20~km/s. Details of recent
measurements of different properties of these satellites of the M31 can
be found in Armandroff et al. (1998) and Grebel (1998). We also
included CAS dSph (Grebel \& Guhathakurta, 1998) in our list with \Vc~
in the range $(10-20)~\kms$. One satellite (AND II) can be formally
included in both lists (MW and M31). It is 271~kpc from the M31, but being
at the distance of 525~kpc from MW it is should also be counted 
as the MW satellite. Since this is the only such case, we have decided to
count it only once -- as a satellite of M31.

\begin{deluxetable}{lllccrcccc}
\tablecolumns{9}
\tablecaption{Parameters of simulations}
\tablehead{\colhead{Model}\hfil & \colhead{$\Omega_0$}\hfil &  \colhead{h}\hfil & \colhead{$\sigma_8$} 
 & \colhead{$m_{\rm particle}$} & \colhead{\nstep} & \colhead{Resolution} & Box & $N_{\rm part}$\\ 
 &  & && (\msunh)& & ($h^{-1}$pc)& ($h^{-1}$Mpc) & &}
\startdata
      SCDM& 1.0 & 0.5 & 1.0  & $2.05\times 10^6$ & 650-40,000  & 150 & 2.5 &$128^3$\nl 
\char'3CDM& 0.3 & 0.7 & 1.0  & $1.66\times 10^7$ & 650-40,000  & 450 & 7.5 &$128^3$\nl 
\enddata
\end{deluxetable}


\section{Cosmological models and simulations}


To estimate the satellite abundances expected in the hierarchical
models, we have run simulations of two representative cosmologies.
Parameters of the models and simulations are given in Table 2,
where $\Omega_0$ is the density parameter of the matter at $z=0$,
$\sigma_8$ is the rms of density fluctuations on $8\Mpch$ scale estimated by
the linear theory at present time using the top-hat filter. Other
parameters given in Table 2 specify the numerical simulations: mass of
dark a matter particle, $m_{\rm particle}$, defines the mass
resolution, number of time-steps at the lowest/highest levels of
resolution, size of the simulation box, and the number of
dark matter particles. Numbers on resolution refer to the size of
the smallest resolution elements (cells) in the simulations.

The simulations have been performed using the Adaptive Refinement Tree (ART)
$N$-body code (Kravtsov, Klypin \& Khokhlov 1997).  The ART code reaches high
force resolution by refining the mesh in all high-density regions with an
automated refinement algorithm. The \LCDM~
simulation used here was used in Kravtsov et al. (1998) and we refer
the reader to that paper for details and tests. Additional tests and
comparisons with a more conventional AP$^3$M code will be presented in
Knebe et al. (1999). The CDM simulation differs
from the {\LCDM} simulations only in the cosmological parameters and
size of the simulation box. Our intent was to use the much more
realistic {\LCDM} model for comparisons with observations, and to use
the CDM model to test whether the predictions depend sensitively on
cosmology and to somewhat broaden the dynamical range of the
simulations. Jumping ahead, we note here that results of the CDM
simulation are close to those of the \LCDM~ simulation as far as the
circular velocity function of satellites is concerned. This indicates
that we are dealing with general prediction of hierarchical scenarios,
not particular details of the \LCDM~ model. Nevertheless, we do expect
that some details of statistics and dynamics of the satellites may depend
on parameters of the cosmological models.

The size of the simulation box is defined by the requirement of high
mass resolution and by the total number of particles used in our
simulations. DM halos can be identified in simulations if they have
more than $\sim 20$ particles (KGKK). Small satellites of the MW and
Andromeda have masses of $\sim (1-5)\times10^8M_{\odot}$. Thus, mass
of a particle in the simulation should be quite small: $\lesssim
10^7M_{\odot}$. Therefore, the number of particles in our simulations
($128^3$) dictates the box size of only a few megaparsec across. This
puts significant constraints on our results. The number of massive
halos, for example, is quite small. In the CDM simulation we have only
three halos with circular velocity larger than $140~\kms$. The number
of massive halos in the \LCDM~ simulation is higher (eight).

The important issue for our study is the reliable identification of
satellite halos. The problems associated with halo identification
within high-density regions are discussed in KGKK. In this study we use
a halo finding algorithm called Bound Density Maxima (BDM; see KGKK and
{\Colin} et al. 1998). The source code and description of the version
of the BDM algorithm used here can be found in Klypin \& Holtzman
(1997).  The main idea of the BDM algorithm is to find positions of
local maxima in the density field smoothed at a certain scale and to
apply physically motivated criteria to test whether the identified site
corresponds to a gravitationally bound halo. The algorithm then
computes various properties and profiles for each of the bound halos
and constructs a uniform halo catalog ready to be used for analysis.
In this study we will use the maximum circular velocity as the halo's 
defining property. This allows us to avoid the problem of
ambiguous mass assignment (see KGKK for discussion) and makes it easier
to compare the results to observations. 

The density maxima are identified using a top-hat filter with radius
$r_s$ (``search radius''). The search is performed starting from a
large number of randomly placed positions (``seeds'') and proceeds by
moving the center of mass within a sphere of radius $r_s$ iteratively
until convergence. In order to make sure that we use a sufficiently
large number of seeds, we used the position of every tenth particle as
a seed. Therefore, the number of seeds by far exceeds the number of
expected halos.  The search radius $r_s$ also defines the minimum
allowed distance between two halos. If the distance between centers of
any of the two halos is $<2r_s$, only one halo (the more massive of the
two) is left in the catalog. A typical value for the search radius is
$(5-10) \hkpc$. We set a lower limit for the number of particles inside
the search radius $N(<r_s)$: halos with $N(<r_s)<6$ are not included in
the catalog. We also exclude halos which have less than 20 bound
particles and exclude halos with circular velocity less than
$10~\kms$. Some halos may have significant substructure
in their cores due, for example, to an incomplete merger. Such cases
appear in the catalogs as multiple (2-3) halos with very similar
properties (mass, velocity, radius) at small separations.  Our strategy
is to count these as a single halo. Specific criteria used to identify
such cases are: (1) distance between halo centers is $\lesssim
30 \hkpc$, (2) their relative velocity in units of the {rms} velocity
of particles in the halos $\Delta v/v$ is less than 0.15, and (3) the
difference in mass is less than factor 1.5. If all the criteria are
satisfied, only the most massive halo was kept in the catalog.

The box size of the simulations clearly puts limitations on sizes and
masses of halos. In a few megaparsec box, one does not find large
groups or filaments. The mean density in the simulation boxes, however, is
equal to the mean density of the Universe, and thus we expect our
simulations to be representative of the field population of galaxies
(galaxies not in the vicinity of massive clusters and groups). The
Local Group and field galaxies are therefore our main targets.
Nevertheless, even in the small boxes used in this paper, the number of
halos is very substantial. We find 1000 -- 2000 halos of different
masses and circular velocities in each simulation. This number is large
enough for a reliable statistical analysis.


\section{Satellites: predictions and observations}


Figure 1 presents the velocity distribution function, defined as the
number of halos in a given circular velocity interval per unit volume,
in two \LCDM~ simulations. The smaller-box simulation is the one that
we use in our further analysis. To estimate whether the halo velocity
function is affected by the small-box size, we compare the small-box
result with results from the larger, $60\hMpc$ box, simulation used in
{\Colin} et al. (1998). The latter followed the evolution of $256^3$
particles and had a mass resolution of $1.1\times 10^9\hMsun$.  In the
small box, the total number of halos with $\Vc> 10~\kms$ and
$\Vc>20~\kms$ is 1716 (1066) for the lowest threshold of 20 bound
particles. The numbers change slightly  if a more stringent limit of 25
particles is assumed: 1556 (1052). In the overlapping range of circular
velocities $V_{\rm circ}=(100-200)~\kms$ the velocity function of the
small box agrees very well with that of the 
 {\pspicture(0.5,-1.5)(13.0,14.0)
\rput[tl]{0}(-1.2,14.0){\epsfxsize=12cm
\epsffile{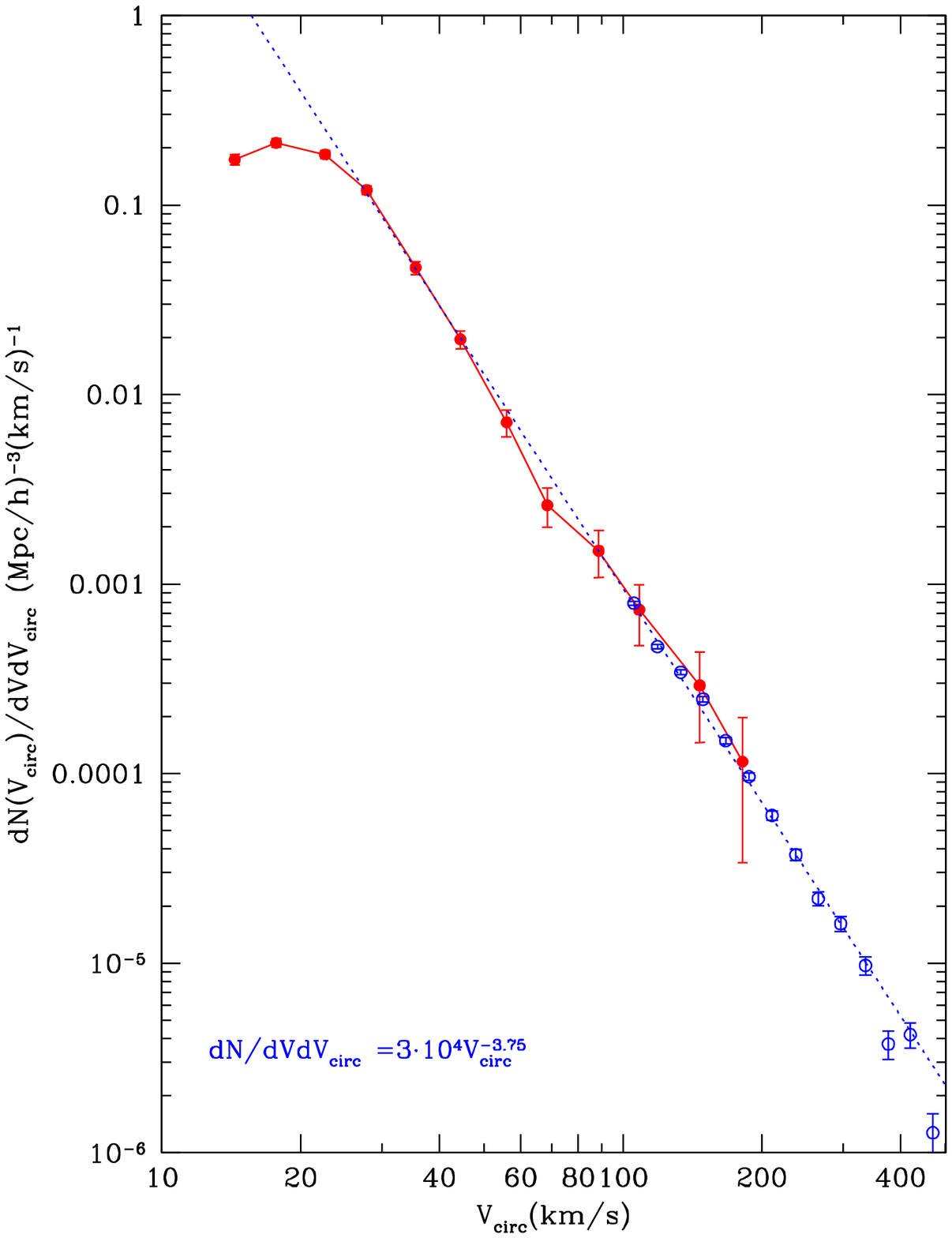}}
\rput[tl]{0}(0.4,2.){
\begin{minipage}{8.7cm}
  \small\parindent=3.5mm {\sc Fig.}~1.--- Differential circular velocity distribution function of dark matter
  halos in the \LCDM~ model. The solid curve and the filled circles
  are results of the small-box (box-size of $7.5\Mpch$)
  simulation. Open circles show the corresponding velocity function in
  larger (box-size of $60\Mpch$) simulation.  Error bars correspond to
  the Poisson noise. The dotted curve is the power-law with the slope
  of -3.75 motivated by the Press-Schechter approximation (see \S 4
  for details).
\end{minipage}
 }
\endpspicture}
large box. This shows that
the lack of long waves in the small-box simulation has not affected the
number of halos with $\Vc < 200~\kms$. 

In the range $V_{\rm circ}\approx 20-400{\ }\kms$ the velocity function
can be accurately approximated by a power law $dN/dVd\Vc \approx
3\times 10^4\Vc^{-3.75}h^3{\rm Mpch}^{-3}/\kms$ motivated by the
Press-Schechter (1974) approximation with assumptions of $M\propto v^3$ and of
the power-law power spectrum with a slope of $n=-2.5$. At higher
circular velocities ($\Vc> 300~\kms$) the fit overpredicts the number
of halos because the above fit neglects the exponential term in the
Press-Schechter approximation. At small \Vc~ ($<20~\kms$) the points
deviate from the fit, which we attribute to the incompleteness of our
halo catalog at these circular velocities due to the limited mass
resolution. Indeed, comparison with the CDM simulation, which has
higher mass resolution, shows that the number of halos increases by
about a factor of three when the threshold for \Vc~ changes from
20~\kms~ to 10~\kms.  We thus estimate the completeness limit of our
simulations to be $\Vc=20~\kms$ for the \LCDM~simulations and
$\Vc=10~\kms$ for the CDM run. Note that for the issue of satellite
abundance discussed below, any incompleteness of the catalogs at these
velocities would increase the discrepancy between observations and
models.

Figure 2 provides a visual example of a system of satellites around a
group of two massive halos in the {\LCDM} simulation. The massive halos
have $\Vc\approx 280{\ }\kms$~ and $\approx 205{\ }\kms$ and masses of
$1.7\times 10^{12}\Msunh$ and $7.9\times 10^{11}\Msunh$ inside the central
$100\hkpc$.  In Figure 3 the more massive halo is shown in more detail.
To some extent the group looks similar to the Local Group though the
distance between the halos is $1.05\Mpch$, which is somewhat larger
than the distance between the MW and M31. Yet, there is a significant
difference from the Local Group in the number of satellites. In
the simulation, there are 281 identified satellites with $\Vc\gtrsim
10~\kms$ within 1.5~\mpch~ sphere shown in Figure 2. The Local Group
contains only about 40 known satellites inside the same radius.

The number of expected satellites is therefore quite large.  Note,
however, that the total fraction of mass bound to the satellites is
rather small: $M_{\rm sat} =0.091\times M_{\rm dm}$, where $ M_{\rm
  dm}=7.8\times 10^{12}\Msunh$ is the total mass inside the sphere.
Most of the mass is bound to the two massive halos. There is another
pair of massive halos in the simulation, which has even more satellites
(340), but the central halo in this case was much larger than M31. Its
circular velocity was $\Vc=302~\kms$.  We will discuss the
correlation of the satellite abundances with the circular velocity of
the host halo below (see Figs. 4 \& 5). The fraction of mass in the
satellites for this system was also small ($\approx 0.055$).

\begin{figure*}[ht]
\pspicture(0,8.0)(15.0,22.6)
\rput[tl]{0}(.0,22.6){\epsfysize=9.cm
\epsffile{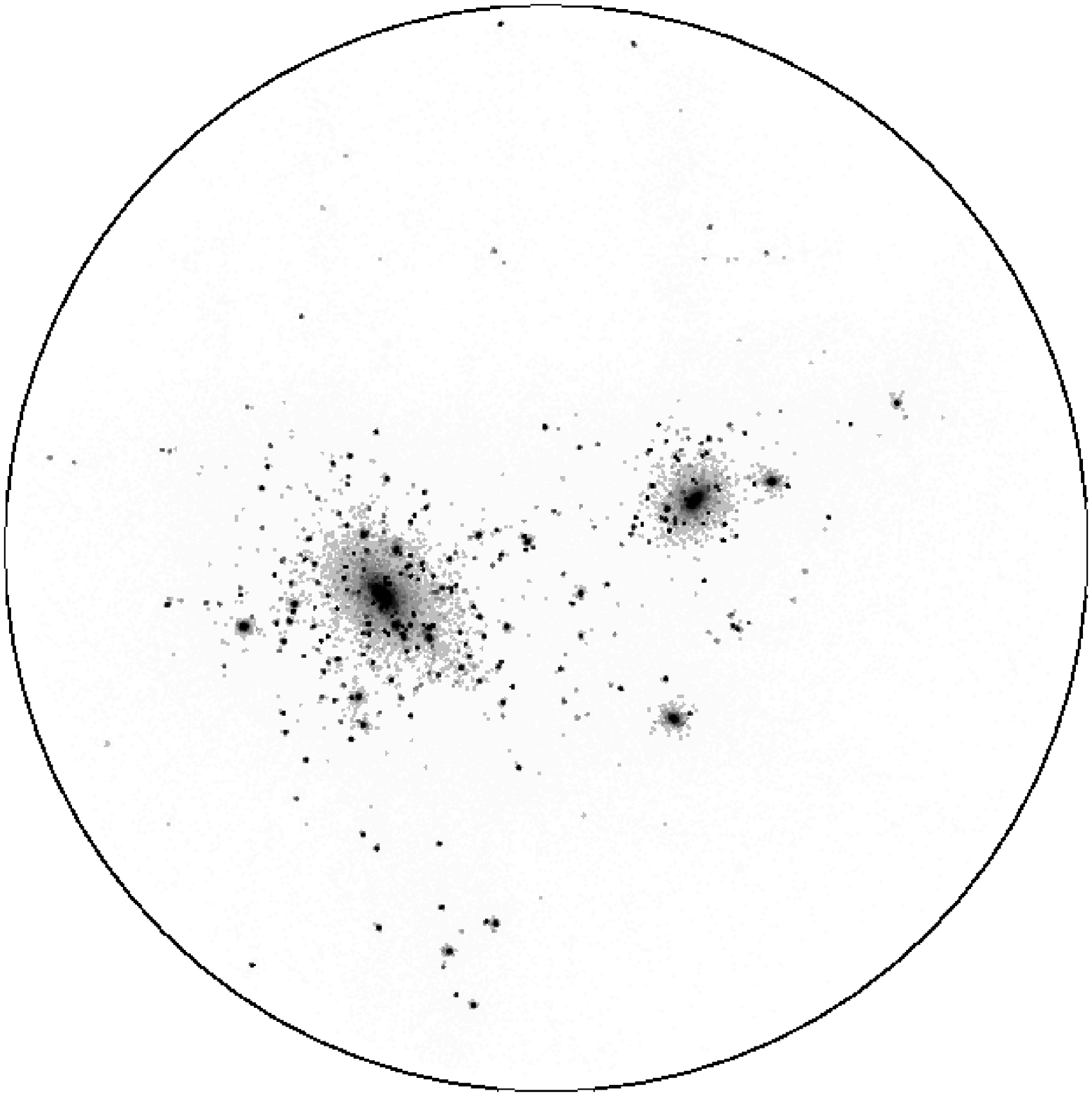}}
\rput[tl]{0}(9.5,22.6){\epsfysize=9.3cm
\epsffile{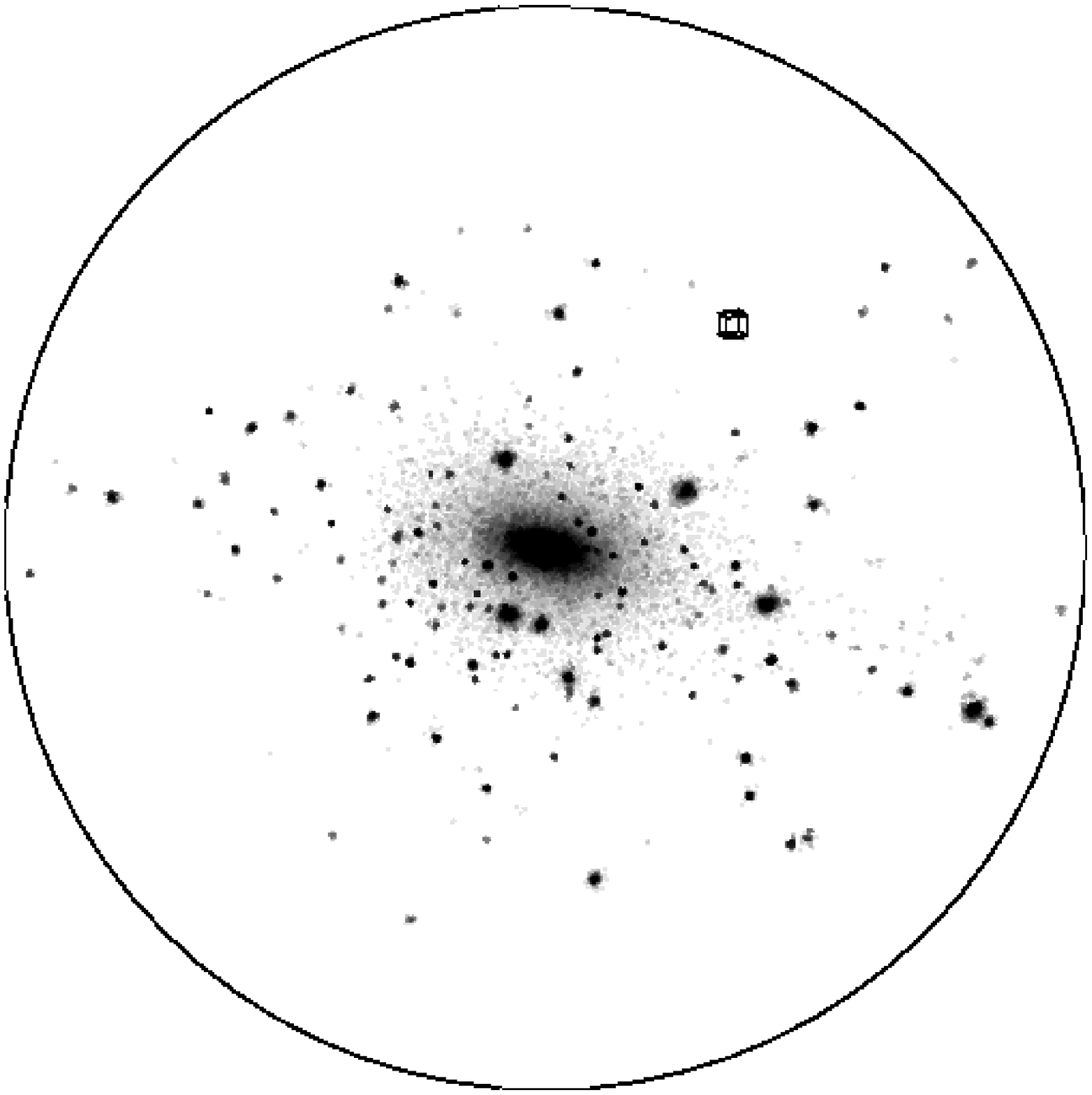}}
\rput[tl]{0}(0.0,12.3){
\begin{minipage}{8.9cm}
  \small\parindent=3.5mm {\sc Fig.}~2.--- Distribution of dark matter
  particles inside a sphere of the radius of 1.5\mpch~(solid circle)
  for a small group of dark matter halos (similar in mass to the Local
  Group) in the \LCDM~ simulation. The group consists of two massive
  halos with circular velocities of $280\kms$ and $205\kms$ (masses of
  $1.7\times 10^{12}\Msunh$ and $7.9\times 10^{11}\Msunh$ inside
  $100\hkpc$ radius) and 281 halos with circular velocities $>10{\
  }\kms$ inside $1.5\mpch$.  The distance between the halos is
  $1.05\Mpch$. To enhance the contrast, we have color-coded DM
  particles on a grey scale according to their local density:
  intensity of each particle is scaled as the logarithm of the
  density, where the density was obtained using top-hat filter with
  $2\hkpc$ radius.
\end{minipage}
}
\rput[tl]{0}(9.6,12.3){
\begin{minipage}{8.9cm}
  \small\parindent=3.5mm {\sc Fig.}~3.--- Distribution of dark matter
  particles inside a sphere of the radius of $0.5\hMpc$ (solid circle)
  centered on the more massive halo shown in Figure 2. The small box
  in the figure has size $20\hkpc$. The color-coding is similar to
  that in Figure 2 except that the local density was obtained using
  top-hat filter of $3\hkpc$ radius.
\end{minipage}
}
\endpspicture
\end{figure*}

Table 3 presents parameters of satellite systems in the \LCDM~
simulation for all central halos with $\Vc>140~\kms$.  The first and
the second columns give the maximum circular velocity \Vc~ and the
virial mass of the central halos. The number of satellites and the
fraction of mass in the satellites are given in the third and fourth
columns. All satellites within $200~(400)~\hkpc$ from the central halos,
possessing more than 20 bound particles, and with $\Vc~ >10~\kms$ were used.
The last two columns give the three-dimensional rms velocity of the
satellites and the average velocity of rotation of the satellite
systems.

Figures 4 and 5 show different characteristics of the satellite systems
in the Local Group (see \S 2), and in the {\LCDM} and the CDM
simulations. Top panels in the plots clearly indicate that the abundance
and dynamics of the satellites depend on the circular velocity (and
thus on mass) of the host halo. More massive halos host more satellites
and the rms velocity of the satellites correlates with host's circular
velocity, as can be expected.  The number of satellites is approximately
proportional to the cube of the circular velocity of the central galaxy
(or halo): $N_{\rm sat} \propto \Vc^3$.  This means that the number of
satellites is proportional to the galaxy mass $N_{\rm sat}\propto M$
because the halo mass is related to \Vc~ as $M\propto \Vc^3$.  The
number of the satellites almost doubles when the distance to the
central halo increases by a factor of two. This is very different from
the Local Group where the number of satellites increases only slightly
with distance.  Note that the fraction of mass in the satellites (see
Table 3) does not correlate with the mass of the central object. The
velocity dispersion decreases with distance, changing by 10\% --
20\% as the radius increases from $200~\hkpc$~ to $400~\hkpc$. We would
like to emphasize that both the number of satellites and the velocity
dispersion have large real fluctuations by a factor of two around their
mean values.

\begin{figure*}[ht]
\pspicture(0,8.0)(15.0,29.)
\rput[tl]{0}(-1.5,29){\epsfysize=13.cm
\epsffile{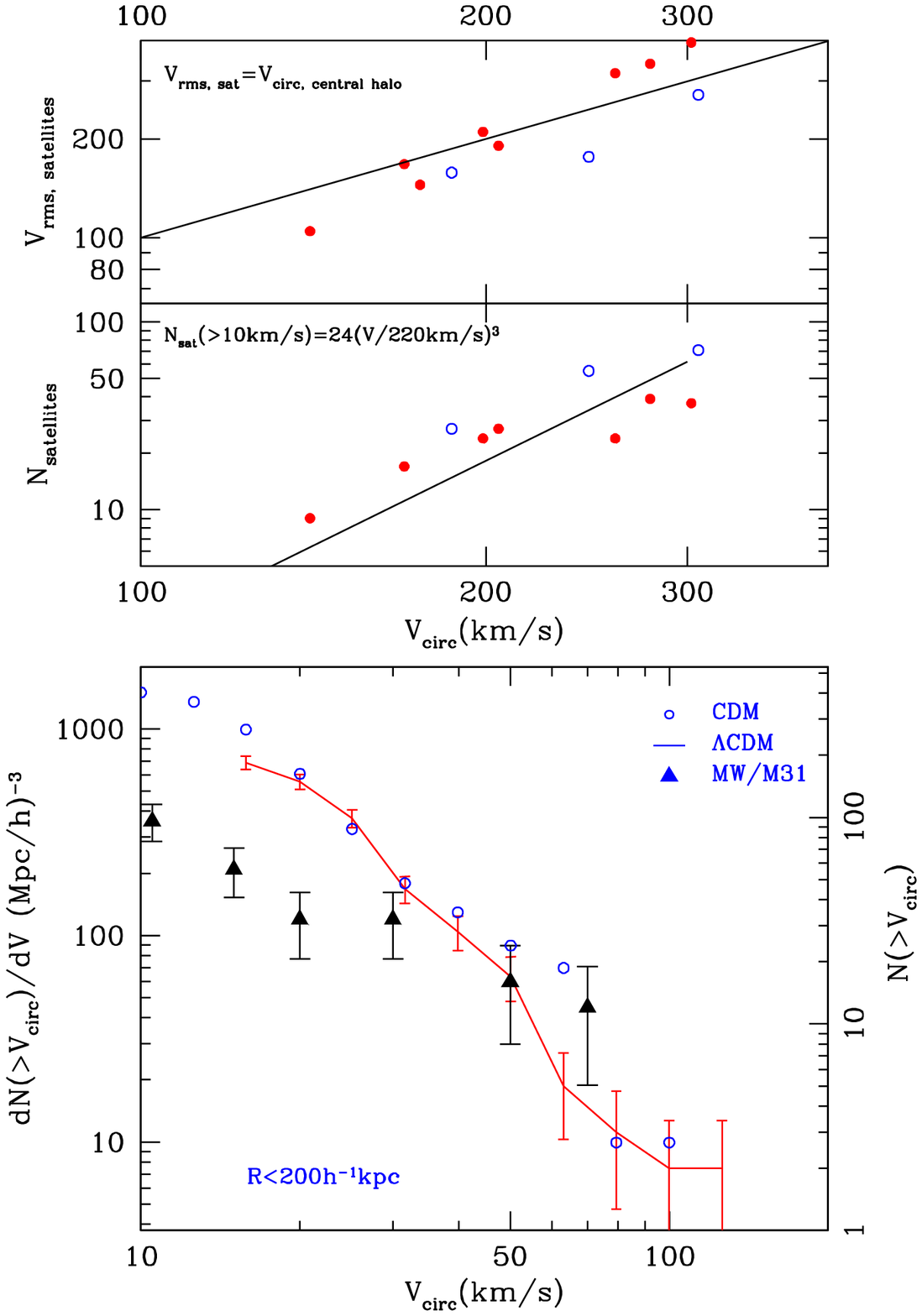}}
\rput[tl]{0}(8.,29){\epsfysize=13.cm
\epsffile{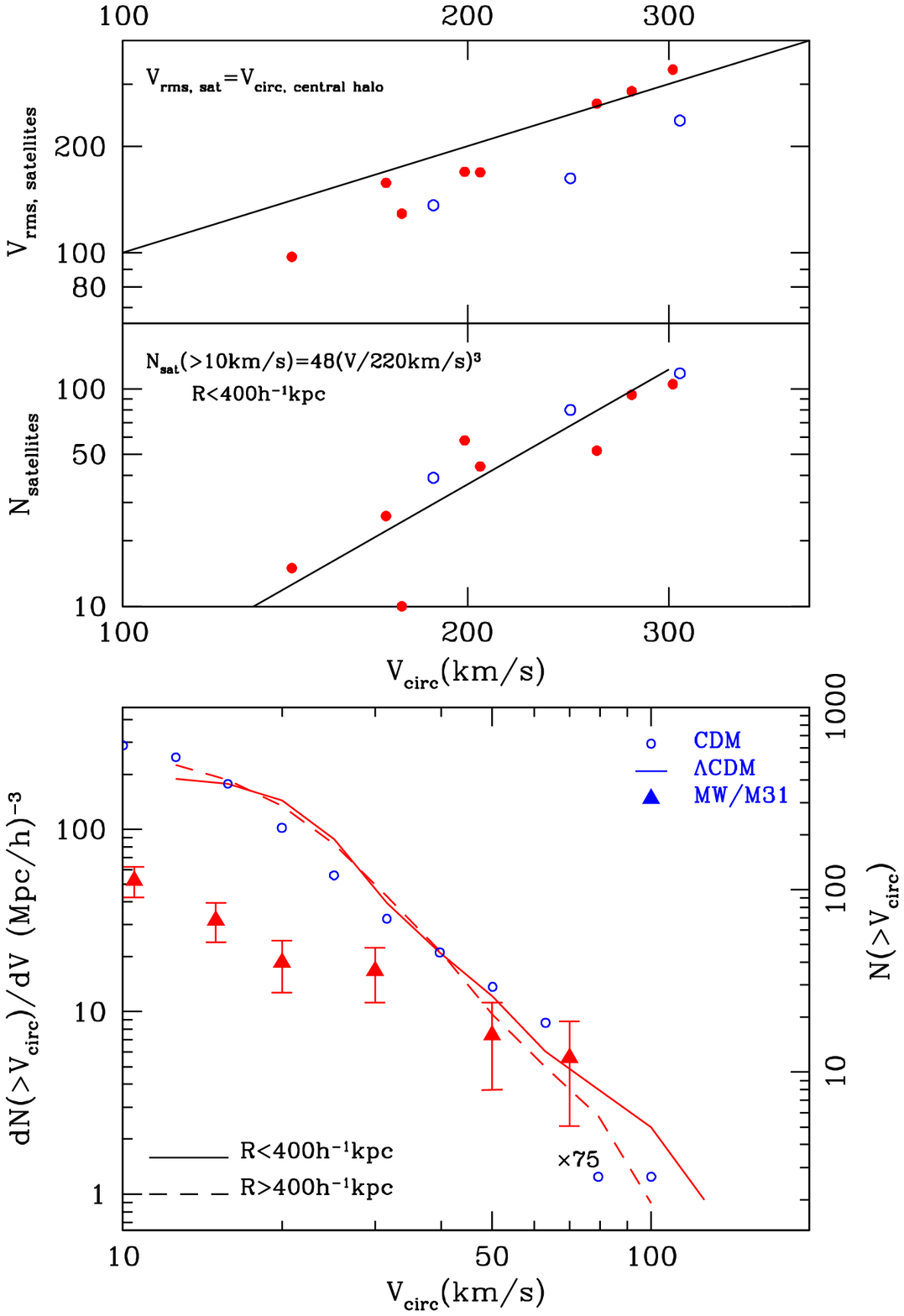}}
\rput[tl]{0}(0.0,15.3){
\begin{minipage}{8.9cm}
  \small\parindent=3.5mm {\sc Fig.}~4.--- Properties of satellite
  systems within $200~\hkpc$ from the host halo.  {\it Top panel:} The
  three dimensional rms velocity dispersion of satellites versus
  maximum circular velocity of the central halo. {\em Solid} and {\em
  open} circles denote {\LCDM} and CDM halos, respectively.  The {\em
  solid line} is the line of equal satellite rms velocity dispersion
  and the circular velocity of the host halo. {\it Middle panel:} The
  number of satellites with circular velocity larger than $10~\kms$
  versus circular velocity of the host halo. The {\em solid line}
  shows a rough approximation presented in the legend. {\it Bottom
  panel:} The cumulative circular velocity distribution (VDF) of
  satellites. {\em Solid triangles} show average VDF of Milky Way and
  Andromeda satellites. {\em Open circles} present results for the CDM
  simulation, while the {\em solid curve} represents the average VDF
  of satellites in the \LCDM~ simulation for halos shown in the upper
  panels. To indicate the statistics, the scale on the right y-axis
  shows the total number of satellite halos in the \LCDM~ simulation.
  Note that while the numbers of massive satellites ($>50~\kms$)
  agrees reasonably well with observed number of satellites in the
  Local Group, models predict about five times more lower mass
  satellites with $V_{\rm circ}< 10-30~\kms$.
\end{minipage}
}
\rput[tl]{0}(9.6,15.3){
\begin{minipage}{8.9cm}
  \small\parindent=3.5mm {\sc Fig.}~5.--- The same as in Figure 4, but
  for satellites within $400~\hkpc$ from the center of a host halo.
  In the bottom panel we also show the cumulative velocity function
  for the field halos (halos outside of $400~\hkpc$~ spheres around
  seven massive halos), arbitrarily scaled up by a factor of 75. The
  difference at large circular velocities $V_{\rm circ}> 50~\kms$~ is
  not statistically significant. Comparison between these two curves
  indicates that the velocity functions of isolated and satellite
  halos are very similar.  As for the satellites within central
  $200\hkpc$ (Figure 4), the number of satellites in the models and in
  the Local Group agree reasonably well for massive satellites with
  $\Vc>50\kms$, but disagree by a factor of ten for low mass
  satellites with $\Vc 10-30~\kms$.
\end{minipage}
}
\endpspicture
\end{figure*}

The bottom panels in Figures 4 and 5 present the {\em cumulative} velocity
distribution function (VDF) of satellites: the number of satellites per
unit volume and per central object with internal circular velocity
larger than a given value of \Vc. Note that the VDF is obtained as the
unweighted average of the functions of individual halos in the interval
$\Vc \approx 150-300{\ }{\kms}$. This was done to improve the
statistics. However, it is easy to check that the amplitude of the VDF
corresponds to the satellite abundance around $\approx 200{\ }{\kms}$
halos. For instance, the average VDF shown in Figure 5 predicts
$\approx 50$ satellites within the radius of $400\hkpc$, while the
upper panel of this figure shows that this is about what we observe for
$\approx 200{\ }{\rm km/s}$ hosts.

\begin{deluxetable}{llclcc}
\tablecolumns{6}
\tablecaption{Satellites in \LCDM~ model inside $R= 200/400\hkpc$ from
central halo}
\tablehead{
\colhead{Halo \vc}\hfil & \colhead{Halo Mass} &\colhead{Number of} &
\colhead{Fraction of mass} &
\colhead{$V_{\rm rms}$} & \colhead{$V_{\rm rotation}$}\\
(km/s) & (\Msunh) &  satellites& in satellites & (km/s) &(km/s) }
\startdata
 140.5 &$2.93\times 10^{11}$&   9/15&0.053/0.112&   99.4/94.4 &  28.6/15.0\nl
 278.2 &$3.90\times 10^{12}$&  39/94&0.041/0.049&  334.9/287.6&  29.8/11.8\nl
 205.2 &$1.22\times 10^{12}$&  27/44&0.025/0.051&  191.7/168.0&  20.0/11.3\nl
 175.2 &$6.26\times 10^{11}$&   5/10&0.105/0.135&  129.1/120.5&  41.5/45.2\nl
 259.5 &$2.74\times 10^{12}$&  24/52&0.017/0.029&  305.0/257.3&  97.1/16.8\nl
 302.3 &$5.12\times 10^{12}$&  37/105&0.055/0.112&  394.6/331.6&  39.4/15.7\nl
 198.9 &$1.33\times 10^{12}$&  24/58&0.048/0.049&  206.1/169.3&  17.7/12.1\nl
 169.8 &$7.91\times 10^{11}$&  17/26&0.053/0.067&  162.8/156.0&   9.3/5.0\nl
\enddata
\end{deluxetable}

The right $y$-axis in the lower panels of Figures 4 and 5 shows the
cumulative number of satellites in all host halos in the \LCDM~
simulation. Error bars in the plots correspond to the Poisson noise.
The dashed curve in Figure 5 shows VDF of all non-satellite halos
(halos located {\em outside} $400~\hkpc$ spheres around the massive
host halos). Comparison clearly indicates that the VDF of the satellite
halos has the same shape as the VDF of the field halos with the only
difference being the amplitude of the satellites' VDF. There are more
satellites in the same volume close to large halos, but the fraction of
large satellites is the same as in the field. We find the same result
for spheres of $200\hkpc$ radius.

The velocity distribution function can be roughly approximated by a
simple power law. For satellites of the Local Group the fit at
$\Vc > 10~\kms$ gives 
\begin{equation}
n(>V)=300\left(\frac{V}{10~\kms}\right)^{-1} (\Mpch)^{-3}, 
\end{equation}
\noindent and
\begin{equation}
n(>V)=45\left(\frac{V}{10~\kms}\right)^{-1} (\Mpch)^{-3},  
\end{equation}

\noindent for $R<200~\hkpc$  and $R<400~\hkpc$, respectively.
 For the \LCDM~ simulation at
$\Vc > 20~\kms$ we obtain:
\begin{equation}
n(>V)=5000\left(\frac{V}{10~\kms}\right)^{-2.75} (\Mpch)^{-3}, 
\end{equation}
\begin{equation}
n(>V)=1200\left(\frac{V}{10~\kms}\right)^{-2.75} (\Mpch)^{-3}, 
\end{equation}
again, for $R<200~\hkpc$ and $R<400~\hkpc$, respectively. This
approximation is formally valid for $\Vc > 20~\kms$, but comparisons
with the higher-resolution CDM simulations indicates that it likely
extends to smaller velocities.  The numbers of observed satellites and
satellite halos cross at around $\Vc=(50-60){\ }\kms$.  This means
that while the abundance of massive satellites ($\Vc > 50{\ }\kms$)
reasonably agrees with what we find in the MW and Andromeda galaxies,
the models predict an abundance of satellites with $V_{\rm circ}>
20~\kms$ that is approximately five times higher than that observed in
the Local Group.  The difference is even larger if we extrapolate our
results to 10~\kms. In this case eq.(4) predicts that on average we
should expect 170 halo satellites inside a $200\hkpc$ sphere, which is
15 times more than the number of satellites of the Milky Way galaxy at
that radius.

\section{Where are the missing satellites?}

Although the discrepancy between observed and predicted satellite
abundances appears to be dramatic, it is too early to conclude that it
indicates a problem for hierarchical models. Several effects can
explain the discrepancy and thus reconcile predictions and
observations. In this section we briefly discuss two possible
explanations: the identification of the missing DM satellites with High
Velocity Clouds observed in the Local Group, and the existence of a large
number of invisible satellites containing a very small amount of luminous
matter either due to early feedback by supernovae or to 
heating of the gas by the intergalactic ionizing background.

\subsection{High Velocity Clouds?}

As was recently discussed by Blitz et al. (1998), abundant
High-Velocity Clouds (HVCs) observed in the Local Group may possibly be
the observational counterparts of the low-mass DM halos unaccounted for by
dwarf satellite galaxies.  It is clear that not all HVCs can
be related or associated with the DM satellites;
there is a number of HVCs with a clear association with the Magellanic
Stream and with the disk of our Galaxy (Wakker \& van Woerden 1997;
Wakker, van Woerden \& Gibson 1999; and references therein).
Nevertheless, there are many HVCs which may well be distant ($>100{\ 
  }{\rm kpc}$; Wakker \& van Woerden 1997; Blitz et al. 1998).
According to Blitz et al., stability arguments suggest diameters and
total masses of these HVCs of $\sim 25{\ }{\rm kpc}$ and $3\times
10^8M_{\odot}$, which is remarkably close to the masses of the
overabundant DM satellites in our simulations.

The number of expected DM satellites is quite high. For the pair of DM
halos presented in Figure 2, we have identified 281 DM satellites with
circular velocities $>10{\ }{\rm km/s}$. Since the halo catalog is
not complete at velocities $\lesssim 20{\ }\kms$ (see \S 4), we expect
that there should be even more DM satellites at the limit
$\Vc=10~\kms$. The correction is significant because about half of the
identified halos have cirlular velocities below $20{\ }\kms$. Using
eq.(3) we predict that the pair should host $(280/2)\times 2^{2.75} =
940$ DM satellites with $\Vc>10~\kms$ within $1.5\hMpc$. A somewhat
smaller number, 640 satellites, follows from eq.(4), if we double
the number of satellites to take into account that we have two massive
halos in the system.

The number of HVCs in the Local Group is known rather poorly. Wakker
\& van Woerden's (1991) all-sky survey, made with 1 degree resolution,
lists approximately 500 HVCs not associated with the Magellanic
Stream.  About 300 HVCs have estimated linewidths (FWHM) of $>20{\
}\kms$ (see Fig.1 in Blitz \etal 1998), the limit corresponding to 3D
rms velocity dispersion of $\sim 15{\ }\kms$.  Stark \etal (1992)
found 1312 clouds in the northern hemisphere, but only 444 of them are
resolved.  The angular resolution of their survey was 2 degrees, but
it had better velocity resolution than the Wakker \& van Woerden
compilation.  Comparisons of low and high resolution observations
indicates that the existing HVC samples are probably affected by
selection effects (Wakker et al. 1999).  The abundance of HVCs thus
depends on one's interpretation of the data.  If we take 1312 HVCs of
Stark \etal (1992), double the number to account for missing HVCs in
the southern hemisphere, we arrive at about 2500 HVCs in the Local
Group. This is more than three times the number of expected DM
satellites. This large number of HVCs also results in a substantial
fraction of mass of the Local Group confined in HVCs. Assuming the
average masses given by Blitz \etal, this naive estimate gives the
total mass in HVCs $7.5\times 10^{11}M_{\odot}$.  If we take mass of
the Local Group to be $\approx 3\times 10^{12}\hMsun$ (Fich \&
Tremaine 1991), the fraction of mass in the HVCs is high:
0.2-0.25. This is substantially higher than the fraction of mass in DM
satellites in our simulations ($\approx 0.05$).

Nevertheless, there is another, more realistic in our opinion, way of
interpreting the data. While it is true that Wakker \& van Woerden
(1991) may have missed many HVCs, it is likely that most of the missed
clouds have small linear size. Thus, the mass should not be doubled
when we make correction for missed HVCs. In this case 500 HVCs (as in
Wakker \& van Woerden sample studied by Blitz \etal) with average dark
matter mass of $3\times 10^8\hMsun$ give in total $1.5\times
10^{11}\hMsun$ or 0.05 of the mass of the Local Group. This is
consistent with the fraction of mass in DM satellites which we find in
our numerical simulations. It should be kept in mind that the small
HVCs may contribute very little to the total mass in the clouds. 

As we have shown above, the number density of DM satellites is a very
strong function of their velocity: $dn(V)/dV\propto V^{-3.75}$.  If the
cloud velocity function is as steep as that of the halos, this might
explain why changes of parameters of different observational samples
produce very large differences in the numbers of HVCs. The mass of a DM
satellite is also a strong function of velocity: $M\propto V^3$. As the
result, the total mass in satellites with velocity {\it less} than $V$
is $\propto V^{2.25}$. The conclusion is that the mass is in the most
massive and rare satellites. If the same is true for the HVCs, we should
not double the mass when we find that a substantial number of small
HVCs were missed in a catalog.

To summarize, it seems plausible that observational data on HVCs are
compatible with a picture where every DM satellite either hosts a dwarf
galaxy (a rare case at small \Vc) or an HVC. This picture relies on the
large distances to the HVCs and can be either confirmed or falsified by
the upcoming observations (Wakker et al. 1999). Note, however, that at
present the observed properties of HVCs (mainly the abundances,
distances, and linewidths) are so uncertain that a more quantitative
comparison is impossible.

\subsection{Dark satellites?}

There are at least two physical processes that have likely operated
during the early stages of galaxy formation and could have resulted in
the existence of a large number of dark (invisible) satellites. The
first process is gas ejection by supernovae-driven winds (e.g., Dekel
\& Silk 1996; Yepes et al. 1997; Mac Low \& Ferrara 1998). This
process assumes at least one initial starformation episode, and thus
should produce some luminous matter inside the host DM
satellites. Indeed, this process may explain the observed properties
of the dwarf spheroidal galaxies in the Local Group (e.g., Dekel \&
Silk 1996; Peterson \& Caldwell 1993; Hirashita, Takeuchi \& Tamura
1998).  It is not clear whether this process can also produce numerous
very low mass-to-light ratio systems missed in the current
observational surveys.  It is likely that some low-luminosity
satellites have still been missed in observations, since several faint
galaxies have been discovered in the Local Group just during the last
few years (see \S 1). What seems unlikely, however, is that
observations have missed so many. This may still be the case if missed
satellites are very faint (almost invisible), but more theoretical
work needs to be done to determine whether gas ejection can produce
numerous very faint systems.  The recent work by Hirashita et al.
(1998) shows that this process may be capable of producing very high
mass-to-light ratio ($M/L$ up to $\sim 1000$) systems of mass
$\lesssim 10^8\hMsun$.

Another possible mechanism is prevention of gas collapse into or
photoevaporation of gas from low-mass systems due to the strong
intergalactic ionizing background (e.g., Rees 1986; Efstathiou 1992;
Thoul \& Weinberg 1996; Quinn, Katz \& Efstathiou 1996; Weinberg,
Hernquist \& Katz 1997; Navarro \& Steinmetz 1997; Barkana \& Loeb
1999). Numerical simulations by Thoul \& Weinberg (1996) and by Quinn
et al. (1996) show that the ionizing background can inhibit gas collapse
into halos with circular velocities $\lesssim 30{ }\kms$.  These
results are in general agreement with more recent simulations by
Weinberg et al. (1997) and Navarro \& Steinmetz (1997). 

As explained by Thoul \& Weinberg, accretion of intergalactic gas
heated by the ionizing background into dwarf $\lesssim 30{ }\kms$
systems is delayed or inhibited because the gas has to overcome
pressure support and is, therefore, much slower to turn around and
collapse. If the collapse may be delayed until relatively late epochs
($z\lesssim 1$), many low-mass DM satellites may have been accreted by
the Local Group without having a chance to accrete gas and form
stars. This would clearly explain the discrepancy between the abundance of
{\em dark matter} halos in our simulations and observed luminous
satellites in the Local Group.  More interestingly, a recent study by
Barkana \& Loeb (1999) shows that gas in small ($\Vc\lesssim 20{\ 
  }\kms$) halos would be photoevaporated during the reionization epoch even
if the gas had a chance to collapse and virialize prior to that.

These results indicate that the ionizing background, of the amplitude
suggested by the lack of the Gunn-Peterson effect in quasar spectra, can
lead to the existence of numerous dark (invisible) clumps of dark matter
orbiting around the Milky Way and other galaxies and thus warrants 
further study of the subject. It would be interesting to explore
potential observational tests for the existence of dark satellites,
given the abundances predicted in hierarchical models. One such
feasible tests, examined recently by Widrow \& Dubinski (1998),
concerns the effects of DM satellites on microlensing statistics in the
Milky Way halo.


\section{Conclusions}


We have presented a study of the abundance and circular velocity
distribution of galactic dark matter satellites in hierarchical models
of structure formation.  Numerical simulations of the \LCDM~ and CDM
models predict that there should be a remarkably large number of dark
matter satellites with circular velocities $\Vc\approx 10-20{\ }\kms$
orbiting our galaxy -- approximately a factor of five more than the
number of satellites actually observed in the vicinity of the Milky Way
or Andromeda (see \S 4). This discrepancy appears to be robust:
effects (numerical or physical) would tend to produce more dark matter
satellites, not less. For example, dissipation in the baryonic
component can only make the halos more stable and increase their chance
to survive. 

Although the discrepancy between the observed and predicted satellite
abundances appears to be dramatic, it is too early to conclude that it
indicates a problem for hierarchical models. Several effects can
explain the discrepancy and thus reconcile the predictions and
observations. If we discard the possibility that $\approx 80\%$ of the
Local Group satellites have been missed in observations, we think that
the discrepancy may be explained by (1) identification of the
overabundant DM satellites with the High Velocity Clouds observed in
the Local Group or by (2) physical processes such as supernovae-driven
winds and gas heating by the ionizing background during the early
stages of galaxy formation (see \S 5). Alternative (1) is
attractive because the sizes, velocity dispersions, and abundance of
the HVCs appear to be consistent with the properties of the
overabundant low-mass halos. These properties of the clouds are deduced
under assumptions that they are located at large ($\gtrsim 100{\ }{\rm
  kpc}$) distances which should be testable in the near future with new
upcoming surveys of the HVCs. Alternative (2) means that the halos
of galaxies in the Local Group (and other galaxies) may contain
substantial substructure in the form of numerous invisible clumps of
dark matter. This second possibility is interesting enough to merit
further detailed study of the above effects on the evolution of gas in
low-mass dark matter halos.

\acknowledgements We are grateful to Jon Holtzman and David Spergel for
comments and discussions. This work was funded by the NSF grant
AST-9319970, the NASA grant NAG-5-3842, and the NATO grant CRG 972148
to the NMSU.  Our numerical simulations were done at the National
Center for Supercomputing Applications (NCSA; Urbana-Champaign,
Illinois).

\end{document}